# Biaxially textured cobalt-doped BaFe$_2$As$_2$ films with high critical current density over 1 MA/cm$^2$ on MgO-buffered metal-tape flexible substrates

Takayoshi Katase[1], Hidenori Hiramatsu[1], Vladimir Matias[2], Chris Sheehan[2], Yoshihiro Ishimaru[3], Toshio Kamiya[1], Keiichi Tanabe[3], and Hideo Hosono[1,4,*]

[1] Materials and Structures Laboratory, Tokyo Institute of Technology, 4259 Nagatsuta-cho, Midori-ku, Yokohama 226-8503, Japan

[2] Superconductivity Technology Center, Los Alamos National Laboratory, Los Alamos, NM 87545, USA

[3] Superconductivity Research Laboratory, International Superconductivity Technology Center, 10-13 Shinonome 1-chome, Koto-ku, Tokyo 135-0062, Japan

[4] Frontier Research Center, Tokyo Institute of Technology, S2-6F East, 4259 Nagatsuta-cho, Midori-ku, Yokohama 226-8503, Japan

**ABSTRACT**

High critical current densities ($J_c$) > 1 MA/cm$^2$ were realized in cobalt-doped BaFe$_2$As$_2$ (BaFe$_2$As$_2$:Co) films on flexible metal substrates with biaxially-textured MgO base-layers fabricated by an ion-beam assisted deposition technique. The BaFe$_2$As$_2$:Co films showed small in-plane crystalline misorientations ($\Delta\phi_{BaFe2As2:Co}$) of ~3$^o$ regardless of twice larger misorientaions of the MgO base-layers ($\Delta\phi_{MgO}$ = 7.3$^o$), and exhibited high self-field $J_c$ up to 3.5 MA/cm$^2$ at 2 K. These values are comparable to that on MgO single crystals and the highest $J_c$ among iron pnictide superconducting tapes and wires ever reported. High in-field $J_c$ suggests the existence of $c$-axis correlated vortex pinning centers.

-----------------------------------------------------------

Footnotes:

[*] Electronic mail: hosono@msl.titech.ac.jp

Iron-pnictide superconductors[1] possess potential for next-generation superconducting tapes or wires because of their relatively high critical temperatures ($T_c$) up to 56 K[2,3] and high upper critical magnetic fields ($H_{c2}//c$) > 100 T with small anisotropy $\gamma (= H_{c2}//ab / H_{c2}//c) = 1–5$ [4–7]. The most important factor for the practical application using high $T_c$ materials including cuprates such as $YBaCu_3O_{7-\delta}$ (YBCO) has been the critical current density ($J_c$) through misaligned grain boundaries (GBs)[8]. The GB properties of the iron pnictides have been examined in cobalt-doped $BaFe_2As_2$ ($BaFe_2As_2$:Co) epitaxial films grown on bicrystal substrates with respect to misorientation angles ($\theta_{GB}$) under high magnetic fields[9] and a self-field[10], because high $J_c > 1$ MA/cm$^2$ have been achieved only in $BaFe_2As_2$:Co[11–13] among thin films of the iron pnictides; these studies have led to recent demonstrations of thin film superconducting devices such as Josephson junctions[14,15] and a SQUID[16]. The $BaFe_2As_2$:Co bicrystal GBs exhibit high self-field $J_c$ unchanged at > 1 MA/cm$^2$ at 4 K up to a large critical GB angle ($\theta_c$) of 9$^o$ [10], which is approximately twice larger than that of YBCO. We confirmed that this difference from YBCO originates from the intrinsic natures of the iron pnictides such as the metallic nature of GB regions that would be carrier-depleted iron pnictides[10].

To date, iron pnictide superconducting wires have been examined mainly by a powder-in-tube technique[17], but their highest $J_c$ are still as low as ~1.0 ×10$^4$ A/cm$^2$ at 4.2 K and the decay slopes with $H$ are large[18]. These deteriorated properties suffer from segregation of secondary phases at GBs, reaction with sheath materials, and/or difficulty in biaxial alignment of the crystallite orientations. Therefore, it is desirable to use ion-beam assisted deposition (IBAD) substrates to prepare biaxially textured films with in-plane misorientation angles $\leq$ 9$^o$ on flexible polycrystalline metal substrates, which have led to high $J_c$ for YBCO coated conductors[19]. The IBAD technique is now widely used to grow biaxially-textured MgO buffer layers with in-plane misorientation angles $\Delta\phi_{MgO} = 6$–$8^o$ [20,21]. For YBCO coated conductors, additional buffer layers such as LaMnO$_3$ and CeO$_2$ are often used to achieve $\Delta\phi \leq 5^o$ to grow high-quality YBCO films[22,23] because YBCO exhibits steep deterioration in $J_c$ with increasing $\Delta\phi_{YBCO} \geq 5^o$ [24], although YBCO coated conductors with $\Delta\phi_{YBCO} = \sim 2^o$ can be grown directly on simpler IBAD-MgO templates if very thick (>1 μm) YBCO films are grown along with a reactive co-evaporation technique[25].

Recently, Iida et al.[26] reported higher $J_c$ (~0.1 MA/cm$^2$ at 8 K) in the epitaxially-grown BaFe$_2$As$_2$:Co films on IBAD-MgO buffered metal substrates with an additional Fe buffer layer; however, the $J_c$ values are still substantially lower than those on MgO single-crystal substrates (~0.5 MA/cm$^2$). Moreover, the ferromagnetic Fe layer would not be appropriate for applications to alternating current devices.

In this letter, we report a high self-field $J_c$ over 1 MA/cm$^2$ at $T \leq 10$ K, which is an order of magnitude higher than those reported to date for iron pnictide tapes[26] and comparable to those of epitaxial films on various single crystals[11–13], in BaFe$_2$As$_2$:Co films fabricated on IBAD-MgO buffered metal-tape substrates (IBAD substrates). An interesting growth mode was observed; i.e., the BaFe$_2$As$_2$:Co films were grown directly on MgO base-layers with large $\Delta\phi_{MgO}$ of 5.5–7.3$^o$, but their $\Delta\phi_{BaFe2As2:Co}$ became much smaller ~3$^o$ even for the small film thickness' of $\leq 220$ nm, which is largely different from the YBCO cases and would be essential for the high $J_c$.

BaFe$_2$As$_2$:Co films were fabricated on IBAD substrates by pulsed laser deposition[12]. The IBAD substrates consisted of top electron-beam-deposited homo-epitaxial MgO layer/IBAD textured MgO layer/amorphous Y$_2$O$_3$ planarizing bed layer/bottom Hastelloy C-276 polycrystalline metal tape. Detailed processes for the IBAD substrates including the planarization process are described in ref. 27. Here, we employed three different IBAD substrates with $\Delta\phi_{MgO}$ = 7.3°, 6.1° and 5.5° ($\Delta\phi_{MgO}$ is evaluated as the in-plane full-width at half maximum (FWHM) value of a MgO 022 rocking curve). We used thin MgO base-layers (homo-epitaxial MgO/IBAD-MgO layers) less than 170 nm, though $\Delta\phi_{MgO}$ < 4° can be obtained if thicker homo-epitaxial MgO layers are grown[21]. All the substrates were cut from a long tape to 1×1 cm$^2$ pieces. The root-mean-square values of the surface roughness of all the homo-epitaxially grown MgO top layers were ~1.5 nm. Each film thickness (130–220 nm) was confirmed by cross-sectional field-emission scanning electron microscopy (FE-SEM). As shown in Fig. 1(a), the thickness of the BaFe$_2$As$_2$:Co layer and the MgO base-layer with $\Delta\phi_{MgO}$ = 6.1° are confirmed to be 150 nm and 50 nm, respectively. Film structures including crystalline quality and crystallites orientation were examined by x-ray diffraction (XRD, anode radiation: CuK$\alpha_1$) at room temperature. Temperature ($T$) dependences of electrical resistivity, $\rho(T)$, and the transport $J_c$ were measured by a four-probe dc method for 300 μm-long, 10 μm-wide micro-bridge patterns formed by photolithography and Ar-ion milling. The $J_c$ was defined from current density–voltage ($J$–$V$) characteristics with a criterion of 1 μV/cm electric field. $H$ dependence of $J_c$ ($J_c(H)$) was evaluated up to $\mu_0 H$ = 9 T applied parallel or perpendicular to the film surface.

      Figures 1(b) shows the out-of-plane XRD patterns for (i) a BaFe$_2$As$_2$:Co epitaxial film on a MgO single crystalline substrate (Film/sc-MgO), (ii) a BaFe$_2$As$_2$:Co film on a IBAD substrate (Film/IBAD) with $\Delta\phi_{MgO}$ = 5.5°, and (iii) a bare IBAD substrate. The out-of-plane XRD pattern of Film/IBAD exhibits intense 00$l$ and weak

110 diffractions assigned to the $BaFe_2As_2$:Co phase. Besides the MgO 002 diffraction peak, a weak diffraction of Fe impurity is also observed, which is similar to the XRD patterns of Film/sc-MgO. The other small peaks originate from the IBAD substrates, and a diffraction of $Y_2O_3$ appears at ~29° because the $Y_2O_3$ layer slightly crystallized during the high temperature growth of the $BaFe_2As_2$:Co films. Inset figure in Fig. 1(b)(ii) shows an out-of-plane rocking curve corresponding to the $BaFe_2As_2$:Co 004 diffraction for Film/IBAD. The FWHM of the rocking curve was $\Delta\omega = 1.5°$, indicating highly $c$-axis orientated films.

In order to evaluate the in-plane orientation relationship between the $BaFe_2As_2$:Co top-layer and the MgO base-layer, pole-figure measurements were performed for the 103 diffractions of $BaFe_2As_2$:Co and the 022 of MgO. Figure 1(c) shows $\beta$ scan results for the $BaFe_2As_2$:Co layer and the MgO base-layer with $\Delta\phi_{MgO} = 5.5°$. Both the patterns exhibit the fourfold rotational symmetry with the same peak angles, which confirms biaxially textured growth of the $BaFe_2As_2$:Co films on the MgO base-layers with the relationship of (001)[100] $BaFe_2As_2$:Co//(001)[100] MgO.

Figure 1(d) shows the $\Delta\phi$ relation between the $BaFe_2As_2$:Co top-layers and the MgO base-layers. The $BaFe_2As_2$:Co layers show small misorientation angles of $\Delta\phi_{BaFe2As2:Co} = 3.2–3.5°$ regardless of the larger $\Delta\phi_{MgO}$ (5.5–7.3°) of the MgO base-layers. Such a significant improvement in the in-plane orientation of the $BaFe_2As_2$:Co top-layer compared with the MgO base-layer may be explained by a "self-epitaxy" effect. It would be noteworthy that the reduction rate in $\Delta\phi_{BaFe2As2:Co}$ with respect to the film thickness is much faster for $BaFe_2As_2$:Co than that observed in YBCO films[28].

Figure 2 shows $\rho(T)$ curves for Film/IBAD and Film/sc-MgO. The onset $T_c$ and the superconducting transition width $\Delta T_c$ for Film/IBAD were 22.0 K and 1.5 K, respectively, and those for Film/sc-MgO were 21.0 K and 1.5 K. The broader transition starting from 26.4 K for Film/IBAD is probably due to inhomogeneous Co content in the film, which would be caused by inhomogeneity in the substrate temperature due to the flexibility of the IBAD substrates. $J$–$V$ characteristics at 2 K for Film/IBAD with $\Delta\phi_{MgO} = 7.3^{o}$, $6.1^{o}$ and $5.5^{o}$ are shown in the inset of Fig. 2. All the $J$–$V$ curves showed sharp superconducting transitions without weak-link behavior at the GBs. Self-field $J_c$ values of 3.6, 1.6 and 1.2 MA/cm$^2$ were obtained for Film/IBAD with $\Delta\phi_{MgO} = 7.3^{o}$, $6.1^{o}$ and $5.5^{o}$, respectively. This result also shows that the difference in $\Delta\phi_{MgO}$ of MgO base-layers did not affect these self-field $J_c$ because the in-plane orientations of the BaFe$_2$As$_2$:Co films are similar ($\Delta\phi_{BaFe2As2:Co} = \sim3^{o}$). The $J_c$ values of the BaFe$_2$As$_2$:Co films on the IBAD substrates are comparable to that on MgO single crystals and remains at $\geq 1$ MA/cm$^2$ at $T \leq 10$ K, which is consistent with the fact that the $\Delta\phi_{BaFe2As2:Co}$ are much smaller than the critical GB misorientation angle ($\theta_c = 9^{o}$) reported in ref. 10.

Figure 3 shows the $J_c(H//c)$ characteristics at $T = 4$ K and 12 K for Film/IBAD with $\Delta\phi_{MgO} = 6.1^{o}$. Those of the Film/sc-MgO[10] are also shown for comparison. All the $J_c(H//c)$ curves monotonically decrease with increasing $H$, but the slopes (d$J_c$/d$H$) for Film/IBAD are smaller than those for Film/sc-MgO, which would originate from a stronger vortex pinning effect along the $c$-axis and also from the higher $T_c$ with high irreversibility field of the Film/IBAD. Figure 3 (b) shows the $J_c(H//ab)$ and $J_c(H//c)$ dependences at 4–18 K for Film/IBAD with $\Delta\phi_{MgO} = 6.1^{o}$. $J_c(H//ab)$ is larger than $J_c(H//c)$ in almost the whole $H$ range at 4 K, while crossovers are observed at 5.0 T at 12 K and 2.9 T at 16 K, respectively. At a higher $T$ of 18 K, $J_c(H//c)$ is greater than $J_c(H//ab)$ in the whole $H$ region. These results are different from the $J_c(H)$ curves for BaFe$_2$As$_2$:Co single crystals, which show the relation of $J_c(H//ab) > J_c(H//c)$ in the whole $H$ and $T$ region[29] that is reasonable based on the intrinsic anisotropy of $H_{c2}$ (i.e.; $H_{c2}//ab > H_{c2}//c$)[6]. These results suggest the existence of strong vortex pinning centers along the $c$-axis in Film/IBAD, which is effective at high $T \geq 12$ K in the low $H$ region less than several T. The inset figure plots the pinning force density $F_p = J_c \times \mu_0 H$ as a function of $H$ measured in $H//c$ at $T = 4$–18 K. Each $F_p$ curve takes a peak ($F_{pmax}$) as indicated by arrows in the inset at $T \geq 12$ K. The largest $F_{pmax}$ is ~8 GN/m$^3$ at 4 K but is three times smaller than those of BaFe$_2$As$_2$:Co epitaxial films on LSAT single crystals (30 GN/m$^3$ at ~12 T)[30]. It suggests that much higher in-field $J_c$ would be realized on IBAD substrates by further optimizing the $c$-axis correlated pinning centers.

In conclusion, we demonstrated high self-field $J_c = 1.2$–3.6 MA/cm$^2$ at 2 K for BaFe$_2$As$_2$:Co films directly grown on IBAD-MgO buffered flexible metal substrates. The biaxially-textured BaFe$_2$As$_2$:Co films exhibited excellent in-plane alignment with $\Delta\phi_{BaFe2As2:Co}$ ~3$^o$ regardless of the larger $\Delta\phi_{MgO} = 5.5$–7.3$^o$ of the MgO base-layers. The in-field $J_c$ of the BaFe$_2$As$_2$:Co films on the IBAD substrates is substantially higher than that for BaFe$_2$As$_2$:Co films on MgO single crystals, probably due to $c$-axis vortex pinning effects. These results imply that high $J_c$ coated conductors can be fabricated with BaFe$_2$As$_2$:Co using less-well-textured templates and with large $\Delta\phi$, which allows for a simple and low-cost process for high-$J_c$, high-$H_c$ superconducting tapes.

## Acknowledgment

This work was supported by the Japan Society for the Promotion of Science (JSPS), Japan, through the "Funding Program for World-Leading Innovative R&D on Science and Technology (FIRST) Program".

**Figure captions**

FIG. 1. (Color online) (a) Cross-sectional FE-SEM image of BaFe$_2$As$_2$:Co film directly grown on IBAD-MgO buffered metal tape substrate with $\Delta\phi_{MgO} = 6.1^o$. (b) Out-of-plane XRD patterns of (i) BaFe$_2$As$_2$:Co epitaxial film grown on MgO single-crystal substrate (Film/sc-MgO), (ii) that on IBAD substrate with $\Delta\phi_{MgO} = 5.5^o$ (Film/IBAD) and (iii) bare IBAD substrate. Inset figure in (ii) shows an out-of-plane rocking curve of the 004 diffraction of a BaFe$_2$As$_2$:Co film on the IBAD substrate and the inset in (iii) illustrates the structure of IBAD substrates. (c) XRD pole-figure $\beta$ scans of the BaFe$_2$As$_2$:Co 103 (top panel) and the MgO 022 diffractions (bottom panel) for Film/IBAD with $\Delta\phi_{MgO} = 5.5^o$. (d) The relationships between $\Delta\phi_{BaFe2As2:Co}$ of the 103 diffraction of the BaFe$_2$As$_2$:Co top and $\Delta\phi_{MgO}$ of the 022 diffraction of the MgO base-layer. The $\Delta\phi_{BaFe2As2:Co}$ of Film/sc-MgO with $\Delta\phi_{MgO} = 0.4^o$ is also shown by a triangle for comparison.

FIG. 2. (Color online) $\rho(T)$ of Film/IBAD (circles) and Film/sc-MgO (triangles). Inset figure shows the $J$–$V$ characteristics at 2 K of Film/IBAD with (left) $\Delta\phi_{MgO} = 7.3^o$, (middle) $6.1^o$ and (right) $5.5^o$.

FIG. 3. (Color online) (a) Magnetic field ($H$) dependences of $J_c$ ($J_c(H//c)$) at 4 K (cricles) and 12 K (triangles) of Film/IBAD with $\Delta\phi_{MgO} = 6.1^o$ (shown by red lines) and Film/sc-MgO (blue lines). (b) $J_c(H)$ measured in $H//ab$ and $H//c$ at temperatures of 4 K, 12 K, 16 K and 18 K indicated by red, orange, green and blue lines, respectively. Closed and open symbols correspond to $J_c(H//ab)$ and $J_c(H//c)$, respectively. Inset shows pinning force density ($F_p$) as a function of $\mu_0H$ applied parallel to the $c$-axis at 4–18 K. The arrows in the inset show the $F_{pmax}$ positions.

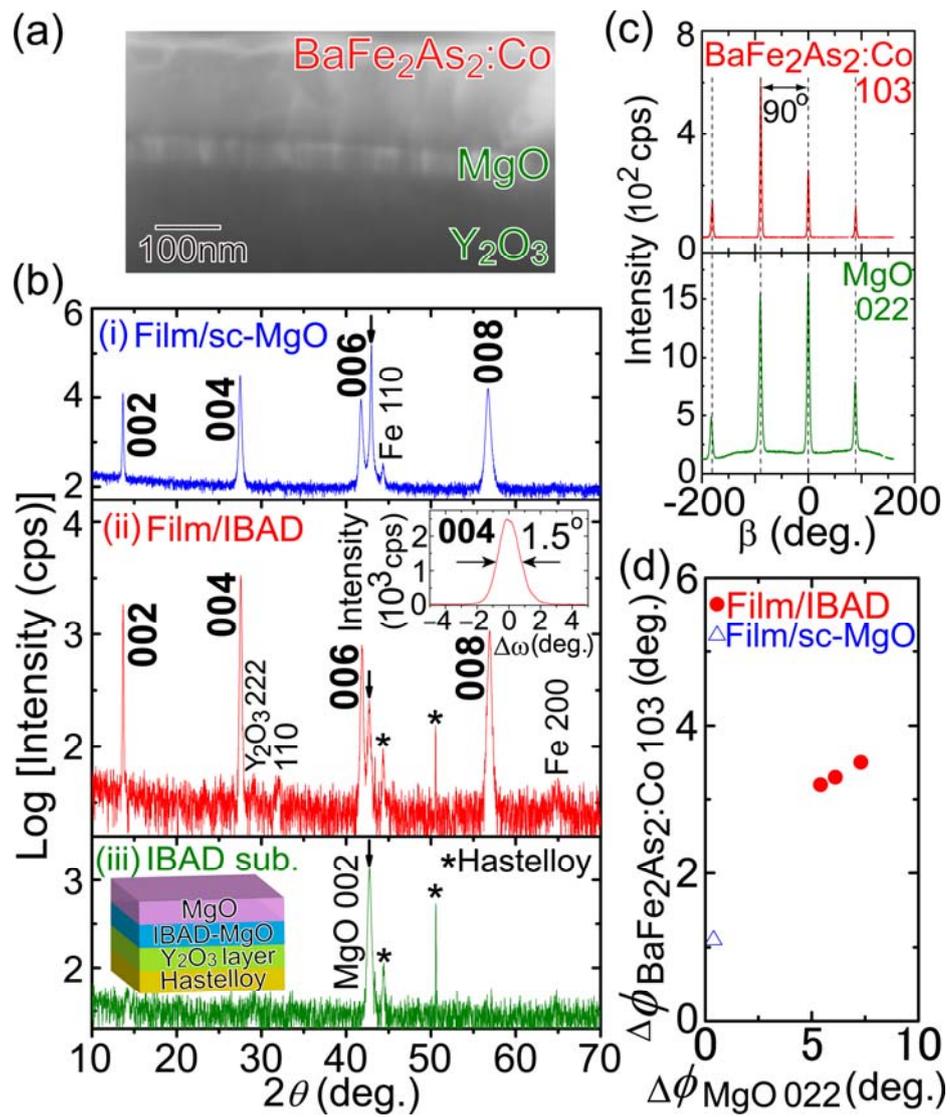

Figure 1

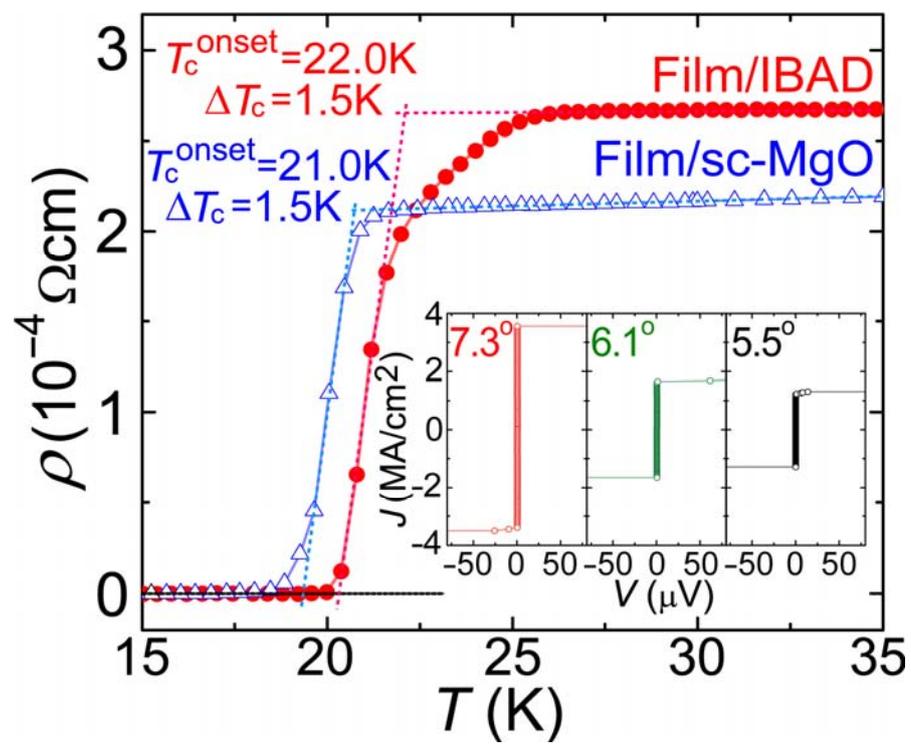

Figure 2

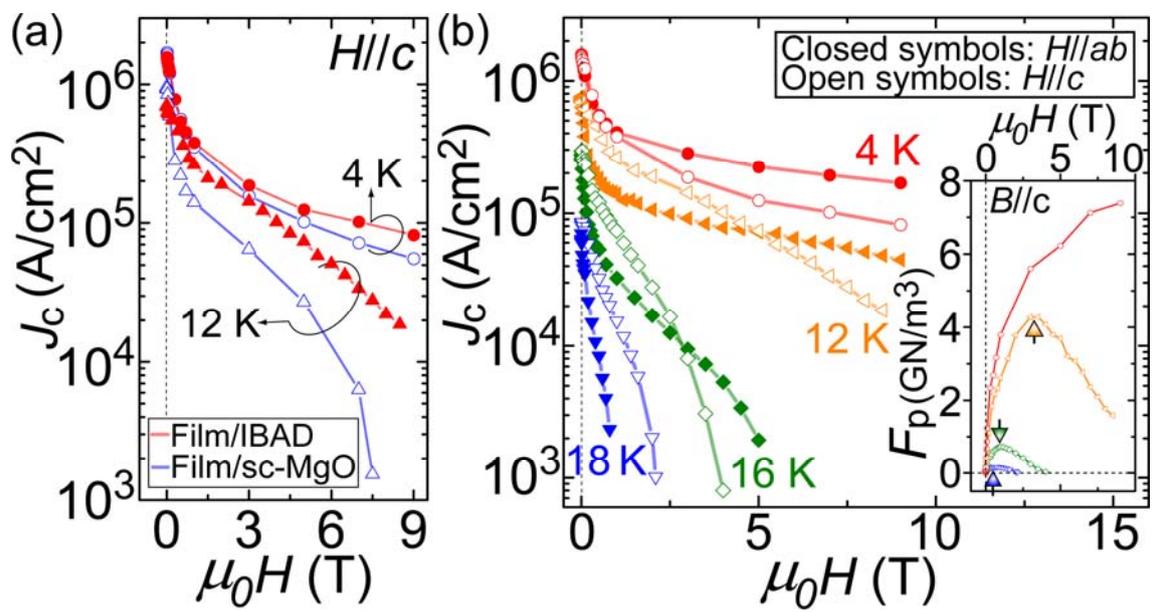

Figure 3